\documentclass[twocolumn,showpacs,preprintnumbers,eqsecnum]{revtex4}
\usepackage{graphicx,dcolumn,bm,amsmath,amssymb}
\def\pt{\widetilde p}
\def\pa{p_A}
\def\pn{p_N}

\begin{document}
\title{\ \ \ \\
       \vspace{-1.7cm} \ \hspace{-6.5cm}
       LYCEN/2002-69, CERN-TH/2002-349, SAGA-HE-193-02 \\
                           \ \\
                           \ \\
        Nuclear modification of \\ transverse-longitudinal 
        structure function ratio}
\author{M. Ericson}
\email{Magda.Ericson@cern.ch}
\address{IPN Lyon, IN2P3-CNRS et UCB Lyon I, 
         F69622 Villeurbanne Cedex, France \\
         and Theory Division, CERN, CH-12111 Geneva, Switzerland}
\author{S. Kumano}
\homepage{http://hs.phys.saga-u.ac.jp}
\email{kumanos@cc.saga-u.ac.jp}
\affiliation{Department of Physics, Saga University,
         Saga, 840-8502, Japan}
\date{November 30, 2002}
\begin{abstract}
\vspace{0.2cm}
We investigate the nuclear effects on the transverse and longitudinal
responses in the deep inelastic region due to the nuclear binding and
nucleon Fermi motion. We display the role of the 
transverse-longitudinal admixture due to the transverse nucleon momentum.
The mixing effect is appreciable at small $Q^2$ values, and gradually
disappears at large $Q^2$ values. The nuclear modification is then
dominated by the binding and Fermi-motion effects which are contained
in the spectral function.
\end{abstract}
\maketitle

\section{Introduction}\label{intro}
\setcounter{equation}{0}

Structure functions in nuclei have been investigated since the discovery
of European Muon Collaboration (EMC) effect \cite{emcsum}. 
Now, as far as the structure functions $F_2^A$ are concerned,
they are relatively well investigated from small to large $x$. 
The $F_2$ structure function contains both longitudinal and transverse
components, so that it is interesting to investigate nuclear effects
on the separated components. 

Experimentally, the nuclear modification of the transverse-longitudinal
ratio was first studied by the HERMES collaboration \cite{hermes00}.
The large effect initially suggested was not confirmed by the neutrino
data \cite{ccfr01} and the HERMES reanalysis \cite{hermes02}.
Barone {\it et al.} \cite{shadow} have studied theoretically
the longitudinal and transverse shadowing. In addition, the role of
the isoscalar meson was investigated by Miller {\it et al.}
\cite{mbk00}.

It is the purpose of this article to discuss, in particular,
the influence of the Fermi motion on the two responses and on their ratio.
At moderate $x$ values, we show that the modification arises from
the admixture of the transverse and longitudinal structure functions.
This effect was introduced by Chanfray {\it et al.} \cite{cdem93}
in the low energy quasi-elastic regime. The mixing arises
from the fact that the photon and nucleon momenta
in the laboratory frame are not aligned. 
We extend here the mixing notion to the deep inelastic regime.
There is also an analogous case in spin physics that $g_1$ and $g_2$
from meson clouds mix \cite{km}. 
In this paper, we show the admixture of transverse
and longitudinal structure functions $F_1$ and $F_L$.
This mechanism produces nuclear modifications of
the transverse-longitudinal ratio $R$ at medium and large $x$.

This paper consists of the following. First, we explain 
the admixture of the transverse and longitudinal
structure functions in a convolution model in Sec. \ref{r}.
Our analysis method is explained in Sec. \ref{analysis},
and numerical results are shown in Sec. \ref{results}.
The results are summarized in Sec.$\,$\ref{sum}.

\vspace{-0.1cm}
\section{Admixture of transverse and longitudinal structure functions}
\label{r}
\setcounter{equation}{0}

\subsection{Structure functions in nuclei}
\label{sfun}
\setcounter{equation}{0}

The cross section for electron scattering from a nucleus is given by
\begin{equation}
\frac{d \sigma}{dE_e^\prime d\Omega_e^\prime} =
            \frac{|\vec p_e^{\, \prime}|}{| \vec p_e |} 
            \, \frac{\alpha^2}{(q^2)^2}
            \, L^{\mu \nu} (p_e, q)
            \, W^A_{\mu \nu} (p_A, q) ,
\label{eqn:cross0}
\end{equation}
where $\alpha$ is the fine structure constant,
$E_e^\prime$ and $\Omega_e^\prime$ are the scattered electron energy
and solid angle, and
$p_e$, $p_e^\prime$, $p_A$, and $q$ are initial electron, final electron,
nucleus, and virtual photon momenta, respectively, in the nuclear
rest frame. In the following discussions, the convention
$-g_{00}=g_{11}=g_{22}=g_{33}=+1$ is used so as to have, for example,
$p_A^2=M_A^2$. The lepton and hadron tensors are denoted as
$L^{\mu \nu}$ and $W^A_{\mu\nu}$, respectively.

The unpolarized hadron tensor is expressed in terms of two structure
functions $W_1$ and $W_2$ by 
parity conservation, time-reversal invariance,
symmetry under the exchange of the Lorentz indices $\mu$ and $\nu$,
and current conservation. The tensors for a nucleus and a nucleon are
expressed as
\begin{align}
W^{A,N}_{\mu\nu} (p_{_{A,N}},  &  q)  =
  - W^{A,N}_1 (p_{_{A,N}}, q) 
  \left ( g_{\mu\nu} - \frac{q_\mu q_\nu}{q^2} \right )
\nonumber \\
& + W^{A,N}_2 (p_{_{A,N}}, q) \, \frac{\pt_{_{A,N} \mu} 
  \, \pt_{_{A,N} \nu}}{p_{_{A,N}}^2} 
\ ,
\label{eqn:hadron}
\end{align}
where $\pt_{\mu} = p_{\mu} -(p \cdot q) \, q_\mu /q^2$.
Projection operators of $W_1^A$ and $W_2^A$ are, then, defined by 
\begin{align}
\widehat P_1^{\, \mu\nu} & = - \frac{1}{2} 
   \left ( g^{\mu\nu} - \frac{\pt_A^{\, \mu} \, 
                 \pt_A^{\, \nu}}{\pt_A^{\, 2}} \right ) ,
\\
\widehat P_2^{\, \mu\nu} & = - \frac{p_A^2}{2\, \pt_A^{\, 2}} 
  \left ( g^{\mu\nu} - \frac{3 \, \pt_A^{\, \mu} \, 
                 \pt_A^{\, \nu}}{\pt_A^{\, 2}} 
  \right ) ,
\label{eqn:p12}
\end{align}
so as to satisfy 
$\widehat P_1^{\, \mu\nu} W^A_{\mu\nu} = W_1^A$ and
$\widehat P_2^{\, \mu\nu} W^A_{\mu\nu} = W_2^A$.

We describe the nuclear tensor $W^A_{\mu\nu}$ by a convolution
model with a spectral function for nucleons in a target nucleus:
\begin{equation}
W^A_{\mu\nu} (\pa, q) = \int d^4 \pn \, S(\pn) \, W^N_{\mu\nu} (\pn, q)
\ ,
\label{eqn:conv}
\end{equation}
where $p_N$ is the nucleon momentum and $S(p_N)$ is the spectral
function which indicates the nucleon momentum distribution
in a nucleus. 
The nuclear structure functions $W_1^A$ and $W_2^A$ are
projected out by operating $\widehat P_1^{\, \mu\nu}$ and
$\widehat P_2^{\, \mu\nu}$ on both sides of Eq. (\ref{eqn:conv}):
\begin{equation}
W_{1,2}^A (\pa, q)   =  \int d^4 \pn \, S(\pn) \, 
           \widehat P_{1,2}^{\, \mu\nu} \, W^N_{\mu\nu} (\pn, q)
\ .
\label{eqn:w12a}
\end{equation}

Because transverse and longitudinal structure functions are
discussed, we specify them in terms of the structure functions
$W_1$ and $W_2$. Photon polarization vectors are given by
$\varepsilon_\pm = \mp (0,1,\pm i, 0)/\sqrt{2}$ and
$\varepsilon_0   = (\sqrt{\nu^2 +Q^2},0,0,\nu)/\sqrt{Q^2}$,
where the photon momentum direction is taken along the $z$ axis
($\vec q / | \vec q | = \hat z$),
$\nu$ is the energy transfer $\nu=E_e-E_e^\prime$,
and $Q^2$ is defined by $Q^2 = -q^2$. 
The structure function for the photon polarization $\lambda$ is
$W^{A,N}_\lambda = \varepsilon_\lambda^{\mu *} \,
                  \varepsilon_\lambda^\nu     \,  W^{A,N}_{\mu\nu}$.
Then, the transverse and longitudinal structure functions are 
given by
\begin{equation}
W^{A,N}_T = \frac{1}{2} \, ( \, W^{A,N}_{+1} + W^{A,N}_{-1} \, ) ,
\ \ \ W^{A,N}_L = W^{A,N}_0 .
\end{equation}
Taking the nucleus or nucleon rest frame, we have
\begin{align}
W^{A,N}_T & = W^{A,N}_1 (p_{_{A,N}}, q) ,
\\
W^{A,N}_L & = (1+\nu_{_{A,N}}^2/Q^2) W^{A,N}_2 (p_{_{A,N}}, q) 
          - W^{A,N}_1 (p_{_{A,N}}, q) ,
\nonumber
\end{align}
where $\nu_A \equiv \nu$, and
the photon momentum in the nucleon rest frame is 
denoted $(\nu_N,\vec q_N)$ with $\nu_N^2 = (p_N \cdot q)^2 /p_N^2$.

Instead of the structure functions $W_1$ and $W_2$, $F_1$ and
$F_2$ are usually used in the analysis. They are defined 
for a nucleus and a nucleon by
\begin{align}
F_1^{A,N} (x_{_{A,N}}, Q^2) & = \sqrt{p_{_{A,N}}^2} 
                  \, W_1^{A,N} (p_{_{A,N}}, q) ,
\\
F_2^{A,N} (x_{_{A,N}}, Q^2) & = \frac{p_{_{A,N}} \cdot q}{\sqrt{p_{_{A,N}}^2}} 
                             \, W_2^{A,N} (p_{_{A,N}}, q) .
\label{eqn:wf12an}
\end{align}
The scaling variables $x_A$ and $x_N$ are given by 
\begin{equation}
x_A = \frac{Q^2}{2 \, p_A \cdot q} = \frac{M_N}{M_A} \, x \ , \ \ \ 
x_N = \frac{Q^2}{2 \, p_N \cdot q} = \frac{x}{z} \ .
\ 
\label{eqn:x}
\end{equation}
Here, the Bjorken scaling variable $x$ and
the momentum fraction $z$ are defined by
\begin{equation}
x   = \frac{Q^2}{2 \, M_N \nu}   \ , \ \ \  
z= \frac{p_N \cdot q}{M_N \, \nu}
\ .
\label{eqn:z}
\end{equation}

The transverse structure functions are $F_1^A$ and $F_1^N$
for a nucleus and a nucleon, respectively.
The longitudinal ones are defined by 
\begin{align}
F_L^{A,N} (x_{_{A,N}}, Q^2) & = \bigg ( 1 + \frac{Q^2}{\nu_{_{A,N}}^2} \bigg ) 
        F_2^{A,N} (x_{_{A,N}}, Q^2)
\nonumber \\
&                    - 2 x_{_{A,N}} F_1^{A,N} (x_{_{A,N}}, Q^2)
\, .
\label{eqn:flla}
\end{align}
The ratio $R_A$ of the longitudinal cross section to the transverse one
is expressed by the function $R_A(x_A,Q^2)$:
\begin{equation}
R_A (x_A, Q^2) = \frac{F_L^A (x_A, Q^2)}{2 \, x_A F_1^A (x_A, Q^2)} 
\ .
\end{equation}
The notations of these structure functions are slightly different
from those of the HERMES collaboration (denoted as H) \cite{hermes00}:
$2 \, x_A \, F_1^A(x_A,Q^2) /A = 2 \, x \, F_1 (x,Q^2)_{H}$,
$F_2^A(x_A,Q^2) / A = F_2 (x,Q^2)_{H}$,
$F_L^A(x_A,Q^2) / A = F_L (x,Q^2)_{H}$, and
$R_A(x_A,Q^2) = R (x,Q^2)_{H}$.
Our structure functions per nucleon are identical to those
of the HERMES collaboration, and the ratio $R_A$ is the same.

\subsection{Transverse-longitudinal admixture}
\label{admix}

Calculating Eq. (\ref{eqn:w12a}) and using
the relations in Eq. (\ref{eqn:wf12an}),
we obtain expressions for $F_1^A$ and $F_2^A$.
Writing these expressions in terms of the transverse and longitudinal
structure functions, we have
\begin{align}
& 2 \, x_A F_1^A (x_A, Q^2) =  \int d^4 \, p_N \, S(p_N) \, z \, 
\frac{M_N}{\sqrt{p_N^2}} 
\nonumber \\
&  \times
\bigg [ \, \bigg ( 1 
            + \frac{\vec p_{N\perp}^{\ 2}}{2 \, \pt_N^{\, 2}} \bigg )
            \, 2 \, x_N F_1^N (x_N, Q^2)
+  \frac{\vec p_{N\perp}^{\ 2}}{2 \, \pt_N^{\, 2}}
            \, F_L^N (x_N, Q^2) \, \bigg ]
\ ,
\label{eqn:trans}
\\
&  F_L^A (x_A, Q^2) =  \int d^4 \, p_N  \, S(p_N) \, z \, 
\frac{M_N}{\sqrt{p_N^2}}
\nonumber \\
&  \times
\bigg [ \, \bigg ( 1 
            + \frac{\vec p_{N\perp}^{\ 2}}{\pt_N^{\, 2}} \bigg )
            \, F_L^N (x_N, Q^2)
+  \frac{\vec p_{N\perp}^{\ 2}}{\pt_N^{\, 2}}
            \, 2 \, x_N F_1^N (x_N, Q^2) \, \bigg ]
\ .
\label{eqn:longi}
\end{align}
In this expression, $\vec p_{N\perp}$ is the nucleon momentum 
component perpendicular to the photon one $\vec q$, and it is 
defined by $\vec p_{N\perp} = \vec p_N - \vec p_{N\parallel}$
with $\vec p_{N\parallel}= (\vec p_N \cdot \vec q) \, \vec q \, 
 /|\vec q \, |^2$. The momentum square $\pt_N^{\, 2}$ is given
by $\pt_N^{\, 2}=p_N^2+Q^2/(4 \, x_N^2)$.

Equation (\ref{eqn:trans}) shows that the {\it longitudinal} nucleonic
structure function enters into the transverse nuclear one and conversely
for the longitudinal nuclear one in Eq. (\ref{eqn:longi}).
The two mixing coefficients are governed, as expected, by the transverse
nucleon momentum, and the admixture terms are proportional to
$O(\vec p_{N\perp}^2/Q^2)$. The mixing is the first cause of
modification of the ratio $R_A$. Moreover, the usual binding 
and Fermi-motion effects, which are implicitly contained
in our convolution expressions, also introduce a nuclear effect.

The original HERMES effect was reported at small $x$.
Their reanalysis indicates that such a large nuclear modification
of $R_A$ does not exist anymore. Our studies are not directly related
to the HERMES finding because the Fermi-motion and binding effects become
important at medium and large $x$. Our predictions are important
for suggesting future experimental efforts for finding such a nuclear
modification of $R_A$.

\section{Analysis}
\label{analysis}

In order to calculate the nuclear structure functions in the previous
section, we need the spectral function for the nucleus. Because
$^{14}$N is one of the nuclei investigated by the HERMES collaboration,
we take this nucleus as an example. 
At this stage, we restrict to the $x$ region, $x<1$, where we may
not need a sophisticated model for the $^{14}$N structure.
The spectral function is simply described by a shell model:
\begin{equation}
S (p_N) = \sum_i | \phi _i (\vec p_N) |^2 \, \delta 
     \bigg ( p_N^0 - M_A + \sqrt{M_{A-i}^2 +\vec p_N^{\, 2}} \, \bigg ) 
\ ,
\label{eqn:sp}
\end{equation}
where $\phi_i$ is the wave function for the $i$-th nucleon,
and $M_{A-i}$ is given by $M_{A-i}=M_A-M_N-\varepsilon_i$
with its single particle energy $\varepsilon_i$.

The nuclear model should be good enough at least to reproduce
gross features of the nucleus, such as nuclear charge radius
and binding energy. As such a model, a density-dependent Hartree-Fock model
by Bonche, Koonin, and Negele (BKN) \cite{bkn} is used in
the following numerical analysis. 
In the used BKN code, the Coulomb interaction is turned off,
so that the single particle energies and wave functions are the same
for the proton and neutron. 
Using calculated wave functions and energies, we obtain
the binding energy and the root-mean-square radius as about $101.7$ MeV
and 2.54 fm. These results agree with the experimental values
about $104.66$ MeV \cite{aw} and 2.45 fm \cite{pb}, so that the model
can be used for our studies of the nuclear structure functions.

Next, we briefly explain the actual integration for evaluating
the transverse and longitudinal structure functions in a nucleus. 
Using the spectral function in  Eq. (\ref{eqn:sp})
and changing the integration variable to $z$ and $p_{N\perp}$,
we have a typical integral of the following form:
\begin{align}
\int d^4 \,  & p_N S(p_N) \, f(x/z,Q^2) 
=  \sum_i \int_x^A dz \int_0^\infty dp_{N\perp}
\nonumber \\
& \times  
\frac{2 \, \pi \, p_{N\perp}}
   { | \partial z / \partial p_{N\parallel} | } \,
 | \phi_i (p_N,\theta_p) |^2   \, f(x/z,Q^2) 
\ ,
\end{align}
where 
$p_N =\sqrt{p_{N\perp}^2+p_{N\parallel}^2}$ and
$\theta_p=\tan^{-1}(p_{N\perp}/p_{N\parallel})$.
In evaluating the integral, the nucleon energy is taken as 
$p_N^0 = M_N + \varepsilon _ i - \vec p_N ^{\, 2} /(2 M _{A-i})$
by using a nonrelativistic approximation for the recoil nucleus.
The momentum fraction $z$ is 
$z= (p_N^0 \nu - p_{N\parallel} \sqrt{\nu^2+Q^2})/(M_N \, \nu)$.
Solving these equations, we can express
$p_N^0$, $p_{N\parallel}$, $\partial z/\partial p_{N\parallel}$
in terms of the integration variables $p_{N\perp}$ and $z$.
In this way, the integrations 
are calculated numerically
with the Hartree-Fock wave functions 
and the following structure functions for a nucleon.

As the leading-order expression of parton distributions, we take
those from the MRST (Martin, Roberts, Stirling, and Thorne)
parametrization in 1998 \cite{mrst}. 
The structure function $F_2$ of the proton is calculated by
$ F_2^p = \sum_i e_i^2 \, x \, ( q_i + \bar q_i ) $.
Isospin symmetry is used for the parton distributions in calculating
$F_2$ of the neutron. 
Since $^{14}$N is isoscalar and the wave functions are identical
for the proton and neutron in the used model, we do not have to take
into account the difference between the proton and the neutron
structure functions.
The nucleon structure function is defined by the average of
the proton and neutron structure functions:
$F_2^N =  ( F_2^p + F_2^n ) /2$.
Using this $F_2^N$ and the SLAC parametrization of 1990 \cite{r1990}
for $R_N$, we calculate the transverse and longitudinal
structure functions by
$F_1^N =  ( 1+Q^2/\nu^2 ) \, F_2^N / [2 \, x \, (1+R_N)]$ and
Eq. (\ref{eqn:flla}).

The transverse and longitudinal structure functions for the nitrogen
nucleus are numerically calculated by Eqs. (\ref{eqn:trans}) and
(\ref{eqn:longi}). Then, longitudinal-transverse ratio $R_A(x,Q^2)$
is evaluated. We should mention that the on-shell structure
functions are used for the bound nucleon, whereas off-shell ones
should be used in principle \cite{offshell}.

\section{Results}
\label{results}

\begin{figure}[h]
\parbox[t]{0.46\textwidth}{
   \begin{center}
\vspace{-0.4cm}
     \includegraphics[width=0.43\textwidth]{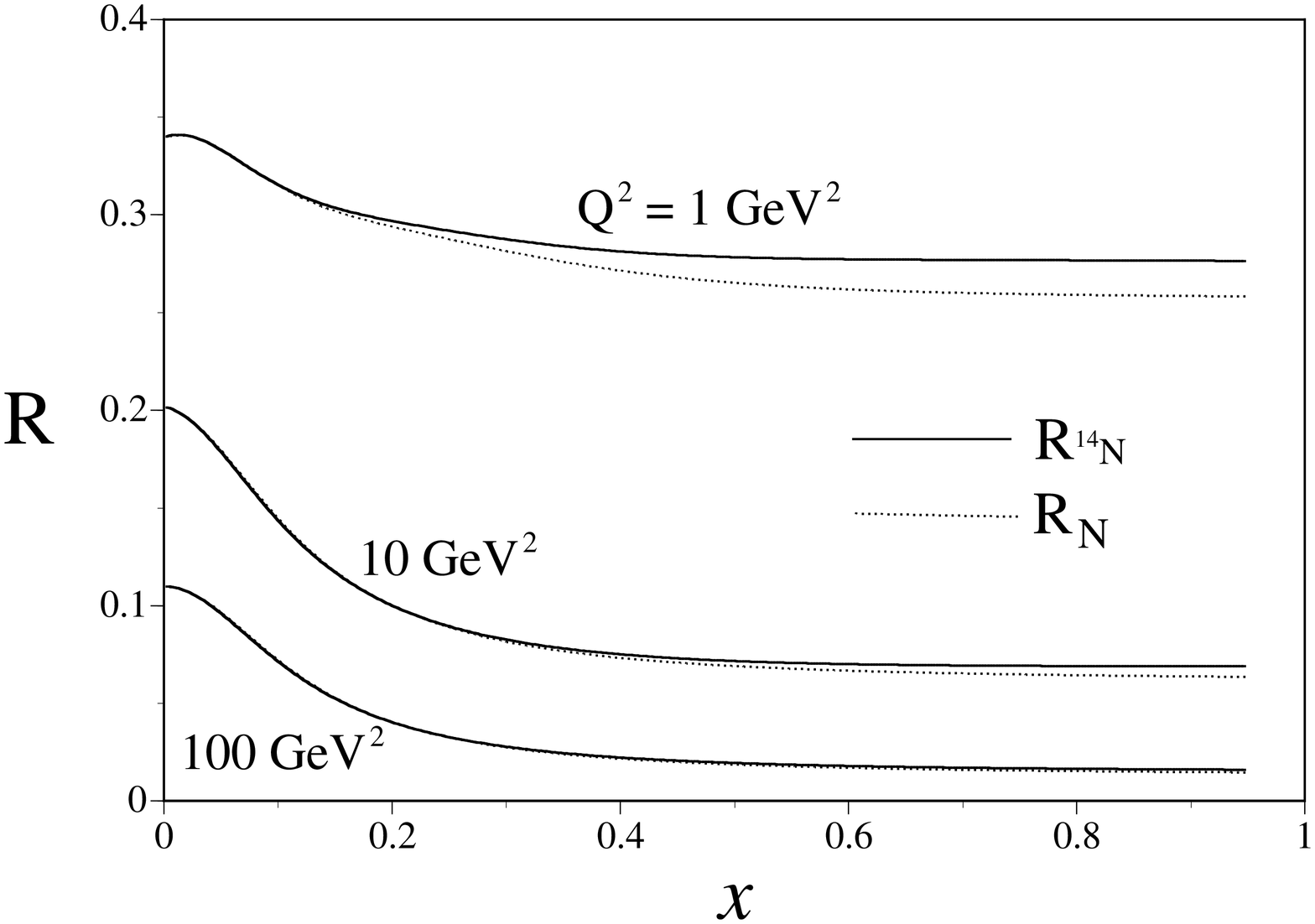}
   \end{center}
   \vspace{-0.9cm}
   \caption{\footnotesize Transverse-longitudinal structure function ratios
               $R(x,Q^2)$ at $Q^2$=1, 10, and 100 GeV$^2$. The solid
               and dotted curves indicate $R_{^{14}N}$ and $R_{N}$,
               respectively.}
   \label{fig:rxna}
}\hfill
\parbox[t]{0.46\textwidth}{
   \vspace{0.3cm}
   \begin{center}
     \includegraphics[width=0.43\textwidth]{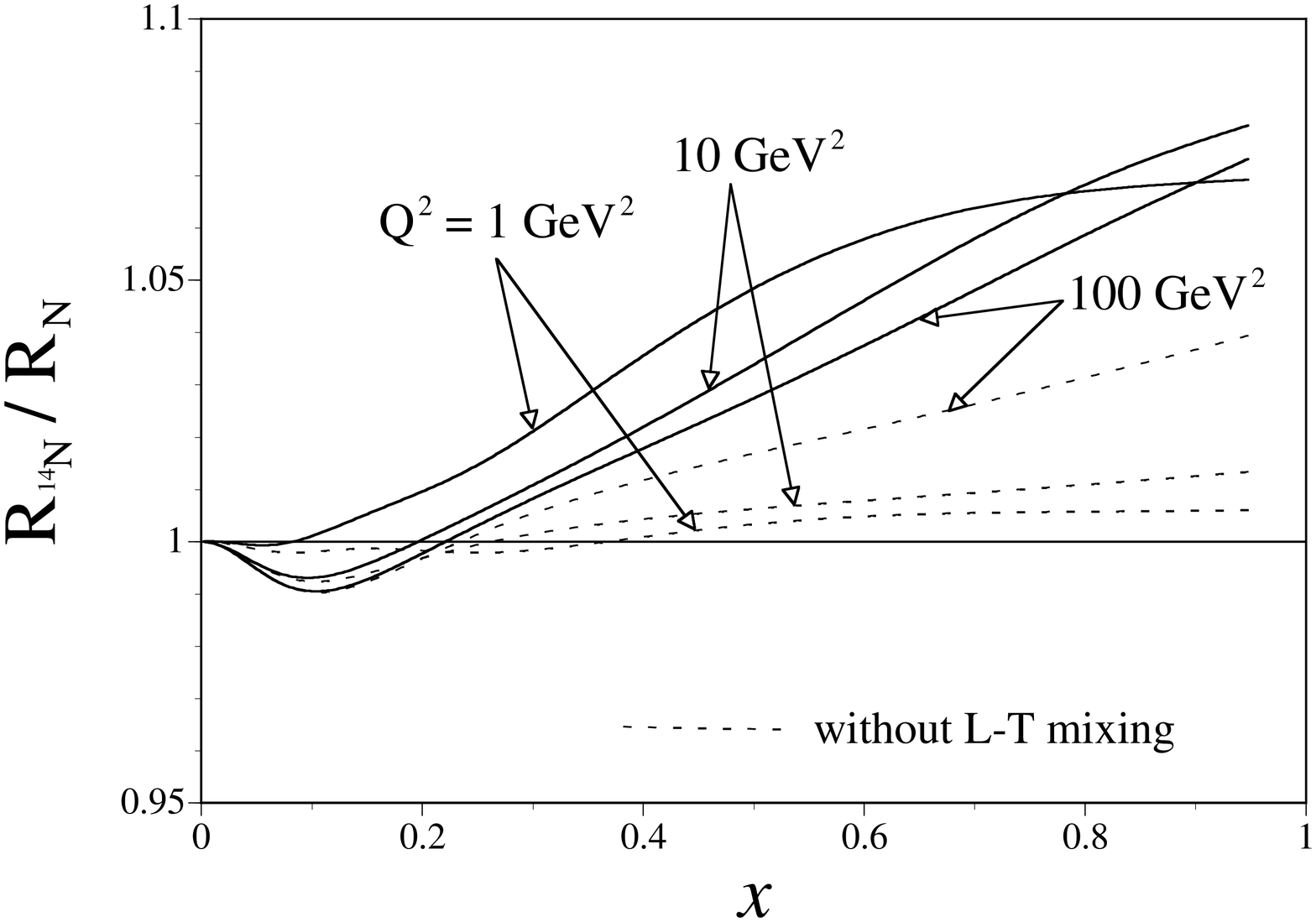}
   \end{center}
 \vspace{-0.9cm}
   \caption{\footnotesize The solid curves show 
           the transverse-longitudinal ratios $R_A(x,Q^2)$
           for $^{14}N$ divided by the corresponding ratio $R_N(x,Q^2)$
           for the nucleon at $Q^2$=1, 10, and 100 GeV$^2$. 
           The dashed curves are obtained without 
           the longitudinal-transverse admixture.}
   \label{fig:rratio}
}
\end{figure}

The obtained $R_{^{14}N}(x,Q^2)$ is displayed
in Fig. \ref{fig:rxna} (solid lines) at $Q^2$=1, 10, and 100 GeV$^2$,
together with the nucleon ratio $R_{N}$ (dotted lines).
As explained in Sec. \ref{admix}, the admixture corrections are
proportional to $O(\vec p_{N\perp}^{\ 2}/Q^2)$, so that they
are large at small $Q^2$ (=1 GeV$^2$) in Fig. \ref{fig:rxna}.
The corrections increase the ratio $R(x,Q^2)$, and they are
conspicuous in the medium and large $x$ regions.
They become small at large $Q^2$ (=10, 100 GeV$^2$); however,
the overall magnitude of $R(x,Q^2)$ also becomes small. 
One should note that the $R_{1990}$ is used also outside the applicable
range ($Q^2<$30 GeV$^2$).

\vspace{-0.42cm}
\noindent
\vspace{-0.0cm}

In Fig. \ref{fig:rratio}, we show the nitrogen-nucleon ratio
$R_{^{14}N}(x,Q^2)$/$R_{N}(x,Q^2)$ at $Q^2$=1, 10, and 100 GeV$^2$.
The solid curves are the ratios with the full nuclear modification.
The dashed ones are obtained by suppressing the $\vec p_{N\perp}^2$
terms in Eqs. (\ref{eqn:trans}) and (\ref{eqn:longi}), 
{\it i.e.} in particular we have suppressed the admixture effect.
Without the admixture, the nuclear modification remains small
at $Q^2$=1 GeV$^2$. However, because of the nucleon transverse motion,
the transverse-longitudinal admixture gives rise to about
5\% positive modification at medium $x$ ($\approx 0.5$).
This 5\% effect is consistent with the value of the mixing coefficient
$\vec p_{N\perp}^2 / \pt_N^{\, 2}
  \approx 4 x_N^2 < \vec p_{N\perp}^{\, 2} >/Q^2 \approx 0.04$.
At larger $Q^2$ (=10 and 100 GeV$^2$), the admixture effects become smaller.
The factor $\vec p_{N\perp}^{\ 2}/Q^2$ becomes small;
however, the ratio $2 x_N F_1/F_L^N$ becomes large
$2 x_N F_1/F_L^N$=14 and 53 for $Q^2$=10 and 100 GeV$^2$, respectively, 
at $x_N=0.5$. Therefore, the overall admixture correction is
of the order of 3 and 1\% for $Q^2$=10 and 100 GeV$^2$, respectively.
The dashed curves in Fig. \ref{fig:rratio} indicate that
the correction from the spectral function also contributes to
the modification. 

\section{Summary}
\label{sum}

Using a convolution description of nuclear structure functions,
we have derived the nuclear modification of the
longitudinal-transverse structure function ratio $R(x,Q^2)$
at medium and large $x$. 
We have found that the conventional convolution description
of the nuclear structure functions leads to the nuclear modification
of the transverse-longitudinal structure function ratio $R(x,Q^2)$.
The physical origin of the modification has been clarified in this paper.
The most important point is the transverse-longitudinal admixture
of the nuclear structure functions 
due to the nucleon momentum transverse to the virtual photon momentum.
The mixing effects we found are moderate at small $Q^2$ and disappear 
at large $Q^2$. The nuclear effects are then dominated by the binding
and Fermi-motion effects contained implicitly in the convolution expression.

In this work, we have used the simplest convolution model.
In particular, we have not included the effects of short-range nucleon
correlation and off-shell structure functions,
and they are expected to be important at large $x$. 
A better description of the nuclear dynamics is necessary to
extend this work to the large $x$ domain.



\end{document}